\documentstyle[12pt]{article}

\def\ar{\rightarrow}
\def\bib{\bibitem}
\def\intx{\int\! d^{\sl 4}x}
\def\intX{\int\! d^{D}X\,}

\def\intP{\int\! \frac{d^{D}P}{(2{\pi})^D}\,}

\def\lar{\longrightarrow}
\def\pa{\partial}
\def\rvec{\!\!\!\!^{^\rightarrow}}

\def\Tr{\,\mbox{Tr}\,}

\def\de{\delta}

\def\la{\lambda}

\def\si{\sigma}

\def\Ga{{\it\Gamma}}
\def\La{{\it\Lambda}}
\def\Om{{\it\Omega}}

\def\Pit{\!{\it\Pi}}

\def\Th{{\it\Theta}}

\def\beq{\begin{equation}}
\def\eeq{\end{equation}}
\def\bed{\begin{displaymath}}
\def\eed{\end{displaymath}}
\def\beqq{\begin{eqnarray}}
\def\eeqq{\end{eqnarray}}
\def\bedd{\begin{eqnarray*}}
\def\eedd{\end{eqnarray*}}

\begin{document}

\centerline{\normalsize\bf CLASSICAL ISOMETRODYNAMICS}

\vspace*{0.9cm}
\centerline{\footnotesize C. WIESENDANGER}
\baselineskip=12pt
\centerline{\footnotesize\it Aurorastr. 24, CH-8032 Zurich}
\centerline{\footnotesize E-mail: christian.wiesendanger@zuerimail.com}

\vspace*{0.9cm}
\baselineskip=13pt
\abstract{A generalization of non-Abelian gauge theories of compact Lie groups is developed by gauging the non-compact group of volume-preserving diffeomorphisms of a $D$-dimensional space ${\bf R\/}^{D}$. This group is represented on the space of fields defined on ${\bf M\/}^{\sl 4}\times {\bf R\/}^{D}$. As usual the gauging requires the introduction of a covariant derivative, a gauge field and a field strength operator. An invariant and minimal gauge field Lagrangian is derived. The classical field dynamics and the conservation laws of the new gauge theory are developed. Finally, the theory's Hamiltonian in the axial gauge and its Hamiltonian field dynamics are derived.}

\normalsize\baselineskip=15pt

\vspace*{0.7cm}
\centerline{\footnotesize \it What is beauty? It is ... the unity of the manifold, the coalescence of the diverse\footnote{Samuel Taylor Coleridge, in {\it On Poesy or Art}.}.}

\section{Introduction}

Is it possible to formulate a consistent new type of perturbatively renormalizable and unitary gauge field theory in four spacetime dimensions? Gauging the infinite-dimensional diffeomorphism group of an "inner" $D$-dimensional space we claim it is.

In four spacetime dimensions there are only a handful of distinct, perturbatively renormalizable and unitary quantum field theoretical models from which to build the theoretical description of the fundamental interactions of Nature. In essence these are models involving scalar bosons, spin-half fermions and the spin-one vector boson theories built from gauging compact Lie groups \cite{stw1}. For one or the other reason attempts to develop a broader number of models within the framework of renormalizable quantum field theory in four spacetime dimensions have failed and yield theories which are either non-renormalizable or violate unitarity.

The class of renormalizable and unitary models mentioned above allows for a successful description of the electromagnetic, weak and strong interactions of Nature with highest precision within the Standard Model (SM) of elementary particle physics \cite{stw2}.

The case of gravity looks rather different. Any attempt at consistently quantizing Einstein's general theory of relativity (GR) or generalizations of GR have failed so far. The quantized theories are either not renormalizable or not unitary \cite{car,clk}. And none of the known gauge theories based on compact Lie groups is a potential candidate for a fundamental description of gravity either.

To overcome this fundamentally unsatisfactory situation many new approaches have been developed - e.g. supersymmetry and superstring theory, just to mention the most important. None of them has, however, provided a consistent model of gravity at the quantum level so far.

In contrast - starting from a well established basis - we propose to generalize the gauge field theory framework from finite-dimensional compact Lie groups to infinite-dimensional gauge groups. Specifically we gauge the non-compact group of the volume-preserving diffeomorphisms of a real space of $D$ "inner" dimensions. In doing so our primary goal is a viable new field theoretical model in itself, a secondary is its potential use in the description of physical interactions, e.g. gravity.

In this paper we develop the corresponding gauge field theory at the classical level. In a second paper the theory is quantized, renormalized at the one-loop level and shown to be asymptotically free. There, also a scetch of the renormalizability to all orders is given \cite{chw3}. Separately we will discuss the potential relevance of the approach for a fundamental description of gravity.

The notations and conventions used follow closely to those used by Steven Weinberg in his classic account on the quantum theory of fields \cite{stw1, stw2}. They are presented in the Appendix.

\section{Gauge Invariance Heuristically Revisited}
In this section we recast the basic concept of gauge invariance \cite{lor} such as to motivate its generalization to infinite-dimensional groups and their representations. We pinpoint the crucial features of finite-dimensional gauge field theories which must continue to hold true in such generalizations to yield properly defined theories.

Let us start with a set of fields $\psi_X (x)$, where for our purposes $X$ is an index - which we aim to take continous later - and $x\in ({\bf M\/}^{\sl 4},\eta)$, the four-dimensional Minkowski spacetime endowed with metric $\eta_{\mu\nu}$.

Next let us act with "matrix" transformations $U_X\,^Y$ on the fields above
\beq \label{1}
\psi_X (x) \lar \psi'_X(x) = U_X\,^Y \psi_Y (x),
\eeq
where we assume the $U_X\,^Y$ to form a group $G$, i.e. for
$U_X\,^Y,\, V_X\,^Y\in G$ we have $U_X\,^Y\, V_Y\,^Z\in G$ and there exists a unit element ${\bf 1}_X\,^Y\in G$ together with an inverse $U^{-1}_{\,X}\,^Y \in G$ fulfilling $U_X\,^Y\, U^{-1}_{\,Y}\,^Z = {\bf 1}_X\,^Z$. The transformations above do not depend on $x$ and "summation" over repeated indices is implied. Note that notions in hyphens such as "matrix" might be generalized beyond their apparent meaning later in this paper.

We now assume that the dynamics of the fields $\psi_X (x)$ shall be specified by a Lagrangian $L_M(\psi_X (x), \pa_\mu\psi_X (x))$ through a variational principle $\de \int\! L_M = 0$. If
\beq \label{2}
L_M(U_X\,^Y \psi_Y, U_X\,^Y \pa_\mu \psi_Y)
= L_M(\psi_X, \pa_\mu\psi_X) 
\eeq 
holds the theory is globally gauge invariant under $G$.

Let us next extend $G$ to a group $U_X\,^Y\ar U_X\,^Y(x)$ of local transformations. Obviously we lose the invariance Eqn.(\ref{2}) as $\pa_\mu$ and $U_X\,^Y(x)$ do not commute anymore. As usual we can compensate for this by the introduction of a "matrix"-valued covariant derivative $(D_\mu)_X\,^Y \equiv (\pa_\mu + A_\mu)_X\,^Y$ with gauge field $A_{\mu\, X}\,^Y$. In order to have local gauge invariance
\beq \label{3}
L_M(U_X\,^Y \psi_Y, D'_\mu U_X\,^Y \psi_Y)
= L_M(\psi_X, D_\mu \psi_X) 
\eeq
we must require $U(x) D_\mu = D'_\mu\, U(x)$ and the gauge field has to transform as
\beq \label{4}
A_{\mu\, X}\,^Y \lar A'_{\mu\, X}\,^Y = U_X\,^Z\, \left(\pa_\mu U^{-1}_{\,Z}\,^Y \right)
+ U_X\,^Z\, A_{\mu\, Z}\,^W U^{-1}_{\,W}\,^Y.
\eeq
Next we define the field strength "matrix" antisymmetric in its spacetime indices 
\beq \label{5}
F_{\mu\nu\, X}\,^Y \equiv [D_\mu, D_\nu]\,_X\,^Y
= \pa_\mu A_{\nu\, X}\,^Y - \pa_\nu A_{\mu\, X}\,^Y +[A_\mu, A_\nu]\,_X\,^Y
\eeq
in the usual way. $F_{\mu\nu\, X}\,^Y$ transforms covariantly
\beq \label{6}
F_{\mu\nu\, X}\,^Y \lar F'_{\mu\nu\, X}\,^Y
= U_X\,^Z\, F_{\mu\nu\, Z}\,^W U^{-1}_{\,W}\,^Y.
\eeq

To specify the dynamics of the $A_{\mu\, X}\,^Y$ we take the well-known gauge-invariant expression of lowest dimension in the gauge fields which formally is 
\beq \label{7}
L_A(A_{\mu\, X}\,^Y, \pa_\nu A_{\mu\, X}\,^Y)
\equiv \Tr F^{\mu\nu}\, F_{\mu\nu}
= F^{\mu\nu}_{\, X}\,^Y F_{\mu\nu\, Y}\,^X, 
\eeq
where we emphasize the word "formally" - meaning that in general Tr has to be shown to be a mathematically well-defined operation. In addition Tr must fulfil the cyclicality property for "matrix" products
\beq \label{8}
\Tr U\,V = \Tr V\,U = U_X\,^Y\, V_Y\,^X 
\eeq
so as to ensure the gauge-invariance of $L_A(A_{\mu\, X}\,^Y, \pa_\nu A_{\mu\, X}\,^Y)$.

Varying $\de \int\! L_A$ w.r.t. $A_{\mu\, X}\,^Y$ yields the equations of motion
\beq \label{9}
[D_\mu, F^{\mu\nu}]_X\,^Y = 0
\eeq
after partial integration and use of the cyclicality of Tr. Here we have assumed that the variations above are unconstrained, i.e. that the $A_{\mu\, X}\,^Y$ are the natural, unconstrained variables of the theory. Note that in the case of compact Lie groups this does only hold true after a further decomposition of the gauge fields w.r.t. to the algebra generators which fully implements the specific constraints coming along with the given Lie algebra.
 
Let us again point to the assumptions behind obtaining a well defined classical theory for the $\psi_X (x)$ combined with the $A_{\mu\, X}\,^Y (x)$. First, there has to be a "vector" space $\psi_X$ on which "matrices" $U$ can act. Second, the $U$ have to form a group under "matrix" multiplication which itself has to be well-defined. Third, there has to exist a Lagrangian which is globally invariant under the above transformation group of the $U$. Fourth, the $U$ can be made dependent on $x$, a covariant derivative $D_{\mu\, X}\,^Y$ and the matrices $A_{\mu\, X}\,^Y$ can be introduced and the gauge fields allow for the transformation law Eqn.(\ref{4}) so as to ensure local gauge invariance. Fifth, the Tr operation in the definition of the gauge field dynamics has to be mathematically well-defined, i.e. finite and cyclical. In quantum theory the Tr must on top be positive-definite so as to allow for a unitary field theory. Finally, the $U_X\,^Y$ obey constraints coming from the very definition of the group they form which translate into corresponding constraints on the $A_{\mu\, X}\,^Y$. These constraints have to be either explicitly solved through a natural choice of independent gauge field variables or they have to be carefully implemented throughout the definition of the theory.

Note that the $X$ label gauge degrees of freedom and that the dependence of $\psi_X (x)$ and $A_{\mu\, X}\,^Y (x)$ on them is not dynamically determined. The true dynamics of the theory is four-dimensional.

The usual application of the framework above - for which all the aforementioned assumptions hold trivially - centers around compact Lie groups. Here the $\psi_X$ form a $M$-dimensional representation space of a $N$-dimensional Lie group. The group elements are represented by $U(\Theta)_X\,^Y =\exp(i\,\Theta^a\,T_a)_X\,^Y $ generated by the $N$ generators $(T_a)_X\,^Y$, $a=1,\dots, N$, and parametrized by $\Theta^a$. Note that the $X=1,\dots, M$ are discrete in this case. 

The $T_a$ form an algebra under commutation $[T_a, T_b]\,_X\,^Y \equiv i\,C_{ab}\,^c\, (T_c)_X\,^Y$, where the $C_{ab}\,^c$ are the structure constants completely characterizing the Lie algebra. Note that the generators $T_a$ fully implement the constraints coming from the definition of the algebra. All the covariant derivative, gauge field and field strength can be decomposed w.r.t. to the $T_a$
\beqq \label{10}
A_{\mu\, X}\,^Y &=& -i\, A^a_\mu \, (T_a)_X\,^Y \nonumber \\
(D_\mu)_X\,^Y &=& \pa_\mu\, \de_X\,^Y - i\, A^a_\mu \, (T_a)_X\,^Y \\
F_{\mu\nu\, X}\,^Y &=& -i\, F^a_{\mu\nu} \, (T_a)_X\,^Y, \nonumber
\eeqq
where $F_{\mu\nu}^a = \pa_\mu A_\nu^a - \pa_\nu A_\mu^a
+ C_{ab}\,^c A_\mu^a\, A_\nu^b$. The $A_\mu^a$ and $F_{\mu\nu}^a$ are the natural, unconstrained variables of the theory.

Finally the Tr operation also decomposes
\beq \label{11}
\Tr F^{\mu\nu}\, F_{\mu\nu}
= - F^{\mu\nu\, a}\, F_{\mu\nu}^b\, (T_a)_X\,^Y (T_b)_Y\,^X 
\eeq
and for a compact Lie group we have
\beq \label{12}
-\, (T_a)_X\,^Y (T_b)_Y\,^X = C(M)\, \de_{ab},
\eeq
where $C(M) > 0$ is the Casimir of the $M$-dimensional representation of the chosen Lie group making the Tr positive definite as required for consistent quantization. This application almost trivially allows to include fields living in representations of the Lie group of different dimensions which is an aspect requiring a separate analysis in the general framework above.

The corresponding classical and quantum field theories have been extensively studied and an appropriate choice of Lie group yields consistent theories for the electromagnetic, weak and strong interactions.

In this paper we want to explore another application taking the $X$ as continous indices. Heuristically the summations now become integrations and the matrices $U_X\,^Y$ kernels $U(X,Y)$ subject to constraints defining the specific group under consideration. Formally it is not difficult to reinterpret all of the above framework in terms of such continous indices $X$. As an example Eqn.(\ref{1}) then translates into
\beq \label{13}
\psi (x,X) \lar \psi'(x,X) = \int dY\, U (X,Y) \psi(x,Y).
\eeq
Note that by the very definition the dependence of $\psi (x,X)$ and $A_\mu (x,X,Y)$ on $X,Y$ is not dynamically determined similar to the Lie group case above. The intrinsic dynamics of the theory remains four-dimensional.

But is there a chance to find groups for which this can become more than a formal playing around with equations and for which this yields a new type of viable gauge theory where the aforementioned assumptions can be shown to hold true - this time for an infinite-dimensional gauge group?

Taking the $X$ as vectors in a $D$-dimensional real flat space and the $U_{F^{-1}}$ as the volume-preserving coordinate transformations of that space $X\ar X'=F(X)$ we claim that the answer is positive. The remainder of this paper is devoted to substantiate our claim\footnote{As we will see the generally non-local kernel $U (X,Y)$ collapses in this case to a localized distribution $U_{F^{-1}} (X,Y) = \de(Y-F^{-1}(X))$. We do not know, however, whether there exist well-defined theories on the basis of groups with truly non-local kernels as well.}.

\section{Diffeomorphism Group Representations and Global Diffeomorphism Invariance}
In this section we analyze representations in field space of the infinite-dimensional volume-preserving diffeomorphism group of ${\bf R\/}^{D}$ and introduce the concept of global diffeomorphism invariance.

Let us start with a $D$-dimensional real vector space ${\bf R\/}^{D}$ which we will call "inner" space in the following. Volume-preserving diffeomorphisms 
\beq \label{14}
X^M\lar X'^N = X'^N(X^M),\:\:M,N=\sl{1,2},\dots,D
\eeq
act as a group ${\overline{DIFF}}\,{\bf R}^D$ under composition on this space. $X'^N (X)$ denotes an invertible and differentiable coordinate transformation of the ${\bf R\/}^{D}$ with unimodular Jacobian
\beq \label{15}
\det\left(\frac{\pa X'^N (X)}{\pa X^M}\right) = 1.
\eeq
The motivation to restrict the analysis to volume-preserving transformations will become clear below.

Next we consider fields $\psi(x,X)$ defined on the product of the four-dimensional Minkowski spacetime $({\bf M\/}^{\sl 4},\eta)$ and the $D$-dimensional "inner" space ${\bf R\/}^{D}$ introduced above. The fields $\psi(x,X)$ are assumed to be infinitely differentiable in both $x$ and $X$ and to vanish at infinity. They form a linear space endowed with the scalar product
\beq \label{16}
\langle \psi \!\mid\! \chi \rangle \equiv \intx\intX \La^D \psi^\dagger (x,X) \cdot\chi(x,X),
\eeq
where we introduce a parameter $\La$ of dimension $[\La]=[X]^{-1}$ so as to define a dimensionless scalar product. $\La$ will play an important role in the definition of the gauge field action later.

Let us now define the representation of ${\overline{DIFF}}\,{\bf R}^D$ in the field space above 
\beqq \label{17}
x^{\nu} &\lar& x'^{\nu} = x^{\nu} \nonumber \\
X^N &\lar& X'^N = X'^N (X) \\
\psi (x,X) &\lar& \psi'(x,X') = \psi (x,X), \nonumber
\eeqq
i.e. the fields transform as scalars under coordinate changes. W.r.t to these transformations the scalar product Eqn.(\ref{16}) is invariant due to the restriction to volume-preserving diffeomorphisms which leave the integration measure invariant.

Note that in addition the fields $\psi(x,X)$ might live in non-trivial representation spaces of both the Lorentz group with spin $s\not= 0$ and of other inner symmetry groups such as $SU(N)$. All these scalar, spinor and gauge vector fields - apart from the gauge vector field related to diffeomorphism invariance to be introduced below - are called "matter" fields in the following. These representations factorize w.r.t translation group representations which is consistent with the Coleman-Mandula theorem.

Let us next assume that the dynamics of the field $\psi(x,X)$ is specified by a Lagrangian of the form 
\beq \label{18}
L_M(\psi, \pa_\mu \psi) = \intX\La^D {\cal L}_M (\psi(x,X), \pa_\mu \psi(x,X)),
\eeq
where the Lagrangian density ${\cal L}_M$ is assumed to be real. The integration measure in "inner" space comes along with a factor of $\La^D$ to keep "inner" integrals dimensionless. The subscript $_M$ denotes generic fermionic and bosonic "matter" in this context. A trivial example is specified by the Lagrangian density
\beq \label{19}
{\cal L}_M(\psi(x,X), \pa_\mu \psi(x,X)) = \frac{1}{2}\,\,\pa^\mu \psi(x,X)\cdot\pa_\mu \psi(x,X) + \frac{1}{2}\,\, m^2 \psi^2(x,X)
\eeq
which describes a free particle of inertial mass $m$ with a continous number of "inner" degrees of freedom in generalization of a free particle with a finite number of degrees of freedom labelled by a discrete index $a$, where $ L_M = \frac{1}{2}\,\,\pa^\mu \psi_a(x)\cdot\pa_\mu \psi^a(x) + \frac{1}{2}\,\, m^2 \psi_a(x)\cdot\psi^a(x)$.

The Lagrangian Eqn.(\ref{18}) is invariant under the gauge transformations Eqns.(\ref{17})
\beqq \label{20}
L_M(\psi', \pa_\mu \psi') &=& \intX'\La^D {\cal L}_M (\psi'(x,X'), \pa_\mu \psi'(x,X')) \nonumber \\
&=& L_M(\psi, \pa_\mu \psi)
\eeqq
due to the unimodularity of the corresponding Jacobian as in Eqn.(\ref{15}). The restriction to volume-preserving or isometric transformations naturally follows from the analysis of the various invariances of the Lagrangian Eqn.(\ref{18}). General coordinate transformation would not leave this free Lagrangian invariant.

Note that the global transformation group ${\overline{DIFF}}\,{\bf R}^D$ is truly infinite-dimensional so that we deal indeed with a generalization of the usual global finite-dimensional Lie group invariance. Note also that the Lagrangian Eqn.(\ref{18}) might be invariant under unitary representations of even bigger symmetry groups acting on fields as in Eqn.(\ref{13}).

To complement our understanding and to prepare for gauging let us finally reformulate the above in an equivalent infinitesimal form starting with the representation of ${\overline{DIFF}}\,{\bf R}^D$ in field space for infinitesimal transformations $X'^N (X) = X^N + {\cal E}^N (X)$
\beqq \label{21}
x^{\nu} &\lar& x'^{\nu} = x^{\nu} \nonumber \\
X^N &\lar& X'^N = X^N \\
\psi (x,X) &\lar& \psi'(x,X) = \psi(x,X) -\, {\cal E}^N (X)\cdot \nabla_N\,\psi(x,X), \nonumber
\eeqq
where we now take a passive view transforming the fields only.

The unimodularity condition Eqn.(\ref{15}) translates into the infinitesimal gauge parameter ${\cal E}^N$ being divergence-free
\beq \label{22}
\nabla_N {\cal E}^N (X) = 0. \nonumber
\eeq
Note the crucial fact that the algebra ${\overline{\bf diff}}\,{\bf R}^D$ of the divergence-free ${\cal E}$s closes under commutation. For $\nabla_N {\cal E}^N = \nabla_N {\cal F}^N = 0$ we have
\beq \label{23} 
\left[{\cal E}^M \cdot \nabla_M, {\cal F}^N \cdot \nabla_N \right] = \left( {\cal E}^M \cdot \nabla_M {\cal F}^N
- {\cal F}^M \cdot \nabla_M {\cal E}^N \right) \nabla_N
\eeq
with
\beq \label{24} 
\nabla_N \left( {\cal E}^M \cdot \nabla_M {\cal F}^N
- {\cal F}^M \cdot \nabla_M {\cal E}^N \right) = 0
\eeq
as required by the finite transformations ${\overline{DIFF}}\,{\bf R}^D$ forming a group under composition.

As a result we can write infinitesimal transformations in field space 
\beq \label{25}
U_{\cal E} (X) \equiv {\bf 1} - {\cal E}(X),\:\: {\cal E}(X)= {\cal E}^N (X)\cdot \nabla_N
\eeq
as anti-unitary operators w.r.t. the scalar product Eqn.(\ref{16}). Both the ${\cal E}(X)$ and the $\nabla_N$ are anti-hermitean w.r.t. the scalar product Eqn.(\ref{16}).

The decomposability of ${\cal E}(X)$ w.r.t. to the operators $\nabla_N$ will be crucial for the further development of the theory, especially for identifying the natural gauge field variables of the theory. 

Introducing the variation $\de_{_{\cal E}} ..\equiv ..' - ..$ of an expression under a gauge transformation we can finally write
\beq \label{26}
\de_{_{\cal E}}\psi (x,X) \equiv \psi' (x,X) - \psi (x,X) = -\, {\cal E}^N (X) \cdot \nabla_N\,\psi (x,X). \eeq
The variation of the Lagrangian density ${\cal L}_M(\psi, \pa_\mu \psi)$ - depending on $x$ and $X$ only through the fields $\psi(x,X)$ and their $x$-derivatives $\pa_\mu \psi(x, X)$ - becomes
\beq \label{27}
\de_{_{\cal E}} {\cal L}_M (\psi, \pa_\mu \psi) = -\, {\cal E}^N \cdot \nabla_N\, {\cal L}_M (\psi, \pa_\mu \psi)
\eeq
implying the global invariance of the corresponding Lagrangian
\beqq \label{28}
\de_{_{\cal E}} L_M &=& \intX \La^D \de_{_{\cal E}} {\cal L}_M (\psi, \pa_\mu \psi) \nonumber \\
&=& - \intX \La^D
\,\nabla_N \cdot \left({\cal E}^N {\cal L}_M (\psi, \pa_\mu \psi)\right) = 0. 
\eeqq
Here we have used $\nabla_N {\cal E}^N = 0$ so that the $\intX$-integration yields zero for fields $\psi$ and gauge parameters ${\cal E}$ vanishing at infinity in $X$-space.

As usual the invariance Eqn.(\ref{28}) implies the existence of $D$ conserved Noether currents
\beqq \label{29} J_M^\nu\,_N &\equiv& \intX \La^D\, \frac{\pa{\cal L}_M}{\pa(\pa_\nu \psi)} \nabla_N\, \psi \nonumber \\
\pa_\nu J_M^\nu\,_N &=& 0,\:\:N=\sl{1,2},\dots,D  \eeqq
and the $D$ time-independent charges
\beq \label{30} {\bf Q}_{M\,N}\equiv \int\! d^{\sl 3}x \, J_M^{\sl 0}\,_N, ,\:\:N=\sl{1,2},\dots,D \eeq
which generate the "inner" global coordinate transformations in field space and will have physical reality in any interpretable theory.

\section{Local Diffeomorphism Invariance, Covariant Derivatives and Gauge Fields}
In this section we introduce local gauge transformations and - to make globally invariant Lagrangians locally invariant - the corresponding covariant derivatives, gauge field and covariant field strength operators. We also define global "inner" scale transformations under which the covariant derivative, gauge field and covariant field strength operators are invariant.

Let us extend the global volume-preserving diffeomorphism group represented in field space to a group of local transformations by allowing ${\cal E}^N (X)$ to vary with $x$ as well, i.e. allowing for $x$-dependent volume-preserving general coordinate transformations ${\cal E}^N (X)\ar {\cal E}^N (x,X)$ in "inner" space. In other words the group we gauge is the group of all "isometric" diffeomorphisms preserving the volume in "inner" space - hence the name {\it Isometrodynamics} for the theory.

In generalization of Eqn.(\ref{25}) we thus consider
\beq \label{31}
U_{\cal E} (x,X) \equiv {\bf 1} - {\cal E}(x,X),\:\: {\cal E}(x,X)= {\cal E}^N (x,X)\cdot \nabla_N.
\eeq
The formulae Eqns.(\ref{21}) together with Eqn.(\ref{22}) still define the representation of the volume-preserving diffeomorphism group in field space.

To assure local gauge covariance for globally diffeomorphism covariant Lagrangian densities as in Eqn.(\ref{27}) we must introduce a covariant derivative $D_\mu$ which is defined by the transformation requirement
\beq \label{32}
D'_\mu \,\, U_{\cal E} (x,X) = U_{\cal E} (x,X)\,\, D_\mu,
\eeq
where $D'_\mu$ denotes the gauge-transformed covariant derivative.

By construction the Lagrangian density in Eqn.(\ref{18}) with covariant derivatives replacing the ordinary ones $\pa_\mu\ar D_\mu$ transforms covariantly under local infinitesimal transformations
\beq \label{33}
\de_{_{\cal E}} {\cal L}_M (\psi, D_\mu \psi)  = - {\cal E}^N (x,X)\cdot \nabla_N \, {\cal L}_M(\psi, D_\mu \psi)
\eeq
and the corresponding Lagrangian is locally gauge invariant
\beqq \label{34}
\de_{_{\cal E}} L_M(\psi, D_\mu \psi)
&=& - \intX \La^D \nabla_N \left({\cal E}^N (x,X)\cdot {\cal L}_M(\psi, D_\mu \psi) \right) \\
&=& 0. \nonumber
\eeqq

Next, to fulfil Eqn.(\ref{32}) we make the usual ansatz
\beq \label{35}
D_\mu (x,X) \equiv \pa_\mu + A_\mu (x,X),\quad A_\mu (x,X) \equiv A_\mu\,^M (x,X)\cdot \nabla_M
\eeq
decomposing $A_\mu (x,X)$ w.r.t the generators $\nabla_M$ of the diffeomorphism algebra in field space. In order to have the gauge fields in the algebra ${\overline{\bf diff}}\,{\bf R}^D$ we impose in addition
\beq \label{36}
\nabla_M A_\mu\,^M (x,X) = 0
\eeq
consistent with $\nabla_N {\cal E}^N (x,X)=0$. As a consequence the usual ordering problem for $A_\mu\,^M$ and $\nabla_M$ in the definition of $D_\mu$ does not arise and $D_\mu$ is anti-hermitean w.r.t to the scalar product defined above.

The requirement Eqn.(\ref{32}) translates into the transformation law for the gauge field
\beq \label{37}
\de_{_{\cal E}} A_\mu (x,X) = \pa_\mu {\cal E} (x,X) -  \left[{\cal E} (x,X), A_\mu (x,X) \right] \eeq
which reads in components
\beq \label{38}
\de_{_{\cal E}} A_\mu\,^M = \pa_\mu {\cal E}^M + A_\mu\,^N \cdot \nabla_N {\cal E}^M - {\cal E}^N \cdot \nabla_N A_\mu\,^M
\eeq
respecting $\nabla_M \de_{_{\cal E}} A_\mu\,^M = 0$. The inhomogenous term $\pa_\mu {\cal E}^M$ assures the desired transformation behaviour of the $D_\mu$, the term $\nabla_N {\cal E}^M \cdot A_\mu\,^N $ "rotates" the "inner" space vector $ A_\mu\,^N $ and the term $-\, {\cal E}^N \cdot \nabla_N A_\mu\,^M$ shifts the coordinates $X^N\ar X'^N = X^N + {\cal E}^N (x,X)$.

Note that the consistent decomposition of both $A_\mu$ and $A'_\mu$ w.r.t. the generators $\nabla_M$ is crucial to the theory's viability. This is ensured by the closure of the algebra Eqn.(\ref{23}) and the gauge invariance of $\nabla_M A_\mu\,^M =0$ for gauge parameters fulfilling $\nabla_N {\cal E}^N =0$. The $A_\mu\,^M$ are the natural variables for the theory.

Let us next define the field strength operator $F_{\mu\nu}$ in the usual way
\beqq \label{39}
F_{\mu\nu}(x,X)&\equiv& \left[D_\mu (x,X), D_\nu (x,X) \right] \nonumber \\
&=& F_{\mu\nu}\,^M (x,X)\cdot \nabla_M
\eeqq
which again can be decomposed consistently w.r.t. $\nabla_M$.
The field strength components $ F_{\mu\nu}\,^M (x,X)$ are calculated to be
\beqq \label{40}
F_{\mu\nu}\,^M (x,X)
&\equiv& \pa_\mu A_\nu\,^M (x,X) - \pa_\nu A_\mu\,^M (x,X) \nonumber \\
&+& A_\mu\,^N (x,X)\cdot \nabla_N A_\nu\,^M (x,X) \\
&-& A_\nu\,^N (x,X)\cdot \nabla_N A_\mu\,^M (x,X) \nonumber \\
&=& D_\mu A_\nu\,^M (x,X) - D_\nu A_\mu\,^M (x,X). \nonumber
\eeqq
Under a local gauge transformation the field strength and its components transform covariantly
\beqq \label{41}
\de_{_{\cal E}} F_{\mu\nu} (x,X) &=& - \left[{\cal E} (x,X) , F_{\mu\nu} (x,X) \right], \nonumber \\
\de_{_{\cal E}} F_{\mu\nu}\,^M &=& F_{\mu\nu}\,^N \cdot \nabla_N {\cal E}^M - {\cal E}^N \cdot \nabla_N F_{\mu\nu}\,^M. 
\eeqq
As required for algebra elements $\nabla_M F_{\mu\nu}\,^M = 0$ and $\nabla_M \de_{_{\cal E}} F_{\mu\nu}\,^M = 0$ for gauge fields fulfilling $\nabla_M A_\mu\,^M = 0$ and gauge parameters fulfilling $\nabla_N {\cal E}^N =0$.

Finally it is useful to give the transformation laws for the gauge field and field strength components under finite active transformations
\beqq \label{42}
x^{\nu} &\lar& x'^{\nu} = x^{\nu} \nonumber \\
X^N &\lar& X'^N = X'^N (x,X) \nonumber \\
A_\mu\,^M (x,X) &\lar& A'_\mu\,^N (x,X') = A_\mu\,^M (x,X)\,
\frac{\pa X'^N (x,X)}{\pa X^M} \\
&-& \pa_\mu X'^N (x,X), \nonumber \\
F_{\mu\nu}\,^M (x,X) &\lar& F'_{\mu\nu}\,^N (x,X') = F_{\mu\nu}\,^M (x,X)\,
\frac{\pa X'^N (x,X)}{\pa X^M}, \nonumber
\eeqq
i.e. the gauge fields and field strength transform as vectors under coordinate changes.

Besides the global and local invariance under "inner" coordinate transformations Eqns.(\ref{21}) the theory has another global invariance in "inner" space - namely scale invariance.
Let us give the respective transformation law for a rescaling with scale parameter $\rho\in {\bf R}^+$
\beqq \label{43}
x^{\nu} &\lar& x'^{\nu} = x^{\nu} \nonumber \\
X^N &\lar& X'^N = \rho X^N \\
A_\mu\,^M (x,X) &\lar& A'_\mu\,^M (x,X') = \rho A_\mu\,^M (x,X), \nonumber \\
F_{\mu\nu}\,^M (x,X) &\lar& F'_{\mu\nu}\,^N (x,X') = \rho F_{\mu\nu}\,^M (x,X).
\nonumber
\eeqq
Under Eqns.(\ref{43}) the operators $D_\mu$, $A_\mu$ and $F_{\mu\nu}$ are invariant which will prove crucial to consistently define the theory below.

\section{The Lagrangian}
In this section we introduce a flat metric in the "inner" space and derive the gauge field Lagrangian minimal in the sense of being gauge-invariant and of lowest possible dimension in the fields.

As heuristically motivated in the first Section we propose the local gauge field Lagrangian to be proportional to $\Tr F^2$ - ensuring gauge invariance and at most second order dependence on the first derivatives of the $A_\mu$-fields which is crucial for a quantization leading to a unitary and renormalizable theory.

To make sense of the formal operation $\Tr$ and to define $\Tr F^2$ properly let us start with the evaluation of the differential operator product
\beq \label{44}
F_{\mu\nu}\, F^{\mu\nu} = F_{\mu\nu}\,^M F^{\mu\nu\,N} {\nabla\rvec}_M {\nabla\rvec}_N + 
\nabla_M \left(F_{\mu\nu}\,^M F^{\mu\nu\,N}\right) {\nabla\rvec}_N,
\eeq
where ${\nabla\rvec}_N$ acts on all fields to its right.

To be able to evaluate the trace in a coordinate system we would like to insert complete systems of $X$- and $P$-vectors
\beq \label{45}
{\bf 1} = \intX \mid \!X\rangle\langle X\!\mid\, , \quad\quad
{\bf 1} = \intP \mid \!P\rangle\langle P\!\mid
\eeq
under the $\Tr$-operation and using $\langle X\!\mid\! P\rangle=\exp(i\, P\cdot X)$. This assumes, however, the existence of Cartesian coordinates and a Euclidean metric in "inner" space and of both co- and contravariant vectors w.r.t. that metric.

So in order to proceed let us endow the "inner" $D$-dimensional real vector space ${\bf R\/}^{D}$ with a metric $g_{MN}(x,X)$ and require that its geometry - which we take as an a priori - is flat, Riem$(g) = 0$. This means that it is always possible to choose Cartesian coordinates globally with the metric $g_{MN}(x,X) = \de_{MN}$ collapsing to the Euclidean metric. Such choices of coordinates amount to partially fixing a gauge and we will call them Euclidean gauges in the following.

Note that under "inner" coordinate transformations the metric transforms as a contravariant tensor
\beq \label{46}
g_{MN}(x,X) \lar g'_{KL}(x,X') = g_{MN}(x,X)\,
\frac{\pa X^M (x,X')}{\pa X'^K}\, \frac{\pa X^N (x,X')}{\pa X'^L}
\eeq
or equivalently
\beq \label{47}
\de_{_{\cal E}} g_{MN}= - {\cal E}^R \cdot \nabla_R g_{MN} - g_{RN}\cdot \nabla_M {\cal E}^R - g_{MR}\cdot \nabla_N {\cal E}^R.
\eeq 

Working in Cartesian coordinates we can insert complete systems of $X$- and $P$-vectors and formally take the trace over the "inner" space
\beqq \label{48}
\Tr \Big\{F_{\mu\nu}\, F^{\mu\nu}\Big\}_\de
&\propto& \intX\intP \langle X\!\mid F_{\mu\nu}\, F^{\mu\nu}
\mid \!P\rangle\langle P\!\mid\! X\rangle \nonumber \\
&=& \intX\intP \Bigg\{ F_{\mu\nu}\,^M\cdot
F^{\mu\nu\,N} \,i P_M \,i P_N \\
&+& \nabla_M \left(F_{\mu\nu}\,^M F^{\mu\nu\,N}\right) i P_N \Bigg\} \nonumber \eeqq
which has to be properly defined. Above we have made use of $\nabla_M\!\mid \!P\rangle = i P_M \!\mid \!P\rangle$ and the subscript $\Tr \{\dots\}_\de$ denotes evaluation in a given coordinate system and for a given metric, in this case Cartesian coordinates and the Euclidean metric. Note that beeing a total divergence in $X$-space and odd in $P$ the second term in Eqn.(\ref{48}) vanishes.

The definition of the remaining $P$-integral requires care in order to avoid potential infinities resulting from the non-compactness of the gauge group. Using $\La$ - introduced above to define a dimensionless "inner" space integration measure - as a cut-off we first calculate
\beq \label{49} 
\int_{\mid P\mid \leq \La}
\frac{d^{D}P }{(2{\pi})^D}\, P_M P_N
= \frac{\Om_D}{D(D+2)} \La^{D+2} \de _{MN},
\eeq
where 
\beq \label{50}
\Om_D\equiv\frac{2\pi^{D/2}}{(2\pi)^D\Ga(D/2)}
\eeq
is the surface of the $D-1$-dimensional sphere up to a factor of $\frac{1}{(2{\pi})^D}$.

Note that regularized in such a way any "inner" $P$-integral over polynomials in $P$ reduces to products of the metric in "inner" space and is as well behaved as the usual sum over structure constants of a compact Lie group is in a Yang-Mills theory.

Using Eqn.(\ref{49}) to evaluate Eqn.(\ref{48}) we now {\it define} a $\La$-dependent trace in Euclidean gauges by
\beqq \label{51}
\Tr_{\La} \Big\{F_{\mu\nu}\, F^{\mu\nu}\Big\}_\de
&\equiv& - \intX \, F_{\mu\nu}\,^M\cdot
F^{\mu\nu\,N} \int_{\mid P\mid \leq \La}
\frac{d^{D}P }{(2{\pi})^D}\, P_M P_N \nonumber \\
&=& - \frac{\Om_D}{D(D+2)} \intX \La^D \, \La F_{\mu\nu}\,^M \cdot \La F^{\mu\nu}\,_M 
\eeqq 
which is easily generalized to arbitrary coordinates
\beq \label{52}
\Tr_\La \Big\{F_{\mu\nu}\, F^{\mu\nu}\Big\}_g = - \frac{\Om_D}{D(D+2)} \intX \, \La^D\, \La F_{\mu\nu}\,^M\cdot \La F^{\mu\nu}\,_M, \nonumber
\eeq
where we have to contract the "inner" indices with $g$ now. The expression above is obviously well defined in any coordinate system and gauge-invariant under the combined transformations of field strenght components Eqns.(\ref{42}) and the metric Eqn.(\ref{46}).

Finally this allows us to write down the Lagrangian for Isometrodynamics 
\beq \label{53}
L_{ID} (A_\nu\,^M, \pa_\mu A_\nu\,^M, \nabla_N A_\nu\,^M, \La)
\equiv \frac{1}{4}\,\frac{D(D+2)}{\Om_D}\,\Tr_\La \Big\{F_{\mu\nu}\, F^{\mu\nu}\Big\}_g
\eeq
and the corresponding Lagrangian density
\beq \label{54}
{\cal L}_{ID} (A_\nu\,^M, \pa_\mu A_\nu\,^M, \nabla_N A_\nu\,^M, \La)
= -\frac{\La^2}{4} \, F_{\mu\nu}\,^M \cdot F^{\mu\nu}\,_M.
\eeq
Both are dimensionless in "inner" space - the Lagrangian density due to the factors of $\La$. The factor of $\frac{1}{4} $ in the definition above has been chosen such as to get the usual normalization of the quadratic part of the Lagrangian density.

Note that the Lagrangian for $\rho\La$ is related to the Lagrangian for a given $\La$ by
\beq \label{55}
L_{ID} (X,A_\nu\,^M(X),\dots, \rho\La) =
L_{ID} (\rho X,\rho A_\nu\,^M(X),\dots, \La)
\eeq
with a similar relation holding for the matter Lagrangian Eqn.(\ref{18}) - the dependence of the theory on $\La$ is controlled by the scale invariance Eqn.(\ref{43}). In other words theories for different $\La$ are equivalent up to "inner" rescalings.

Why have we not simply written down Eqn.(\ref{53})? First, the calculation starting with the $\Tr$-operation shows that the dimensionful parameter $\La$ automatically emerges in the definition of the Lagrangian and that the theory at $\La$ is related in a simple way to the one at $\rho\La$. We would not have uncovered this somewhat hidden, but crucial fact in simply writing down the Lagrangian. Second, we will have to show in the quantized version of the theory that the "kinematic" integrals generalizing the "kinematic" sums over gauge degrees of freedom can be consistently defined. The definition of $\Tr F^2$ is a first example of how this will be achieved.

\section{Lagrangian Field Dynamics}
In this section we develop Lagrangian Isometrodynamics determining the classical field equations which will not depend on the metric $g$ and the most important conservation laws for the theory. 

Note that by definition we always work with fields living in the algebra ${\overline{\bf diff}}\,{\bf R}^D$ from now on. We start with the action for Isometrodynamics
\beq \label{56}
S_{ID} = \frac{1}{4}\,\frac{D(D+2)}{\Om_D}\,\intx \,\Tr_\La \Big\{F_{\mu\nu}\, F^{\mu\nu}\Big\}_g.
\eeq
Variation of Eqn.(\ref{56}) w.r.t. $A^{\nu\,N}$ to get the stationary point 
\beqq \label{57}
\de S_{ID} &=& \frac{D(D+2)}{\Om_D}\,\intx \,\Tr_\La \Bigg\{ \Big(
-\pa^\mu F_{\mu\nu}\,^M {\nabla\rvec}_M \nonumber \\
&+& F_{\mu\nu}\,^R {\nabla\rvec}_R A^{\mu\,M} {\nabla\rvec}_M
- A^{\mu\,R} {\nabla\rvec}_R F_{\mu\nu}\,^M {\nabla\rvec}_M \Big)
\, \de A^{\nu\,N} {\nabla\rvec}_N 
\Bigg\}_g \nonumber \\
&=& \frac{D(D+2)}{\Om_D}\,\intx \,\Tr_\La \Bigg\{\Big(
-\pa^\mu F_{\mu\nu}\,^M \\
&-& A^{\mu\,R}\cdot \nabla_R F_{\mu\nu}\,^M
+ F_{\mu\nu}\,^R\cdot \nabla_R A^{\mu\,M} \Big) \, \de A^{\nu\,N} 
{\nabla\rvec}_N {\nabla\rvec}_M \Bigg\}_g \nonumber \\
&=& 0 \nonumber
\eeqq
yields the field equations
\beq \label{58}
\pa^\mu F_{\mu\nu}\,^M + A^{\mu\,N}\cdot \nabla_N F_{\mu\nu}\,^M
- F_{\mu\nu}\,^N\cdot \nabla_N A^{\mu\,M} = 0
\eeq
which by inspection do not depend on the metric. This means that the metric $g$ is not an independent dynamical field and irrelevant for the dynamics of the gauge fields. Above we have used the cyclicality of the trace, partially integrated and brought all the ${\nabla\rvec}_N$ to the right. Note that under the trace all terms with an odd number of ${\nabla\rvec}_N$ vanish.

The equations of motion can be brought in a covariant form
\beq \label{59}
{\cal D}^{\mu\, M}\,_N F_{\mu\nu}\,^N = 0
\eeq
introducing the covariant derivative ${\cal D}^{\mu\, M}\,_N$ acting on vectors in "inner" space
\beq \label{60}
{\cal D}^{\mu\, M}\,_N \equiv \pa^\mu\, \de^M\,_N + A^{\mu\, L} \cdot \nabla_L\, \de^M\,_N - \nabla_N A^{\mu\,^M}.
\eeq 

By inspection the covariant derivative Eqn.(\ref{60}) respects the gauge algebra and is an endomorphism of ${\overline{\bf diff}}\,{\bf R}^D$ because
\beq \label{61}
\nabla_M {\cal D}^{\mu\, M}\,_N G^N = 0
\eeq
for $\nabla_N G^N = 0$.

Finally we can recast the field equations in coordinate-independent and manifestly covariant form
\beq \label{62}
\left[D_\mu, F^{\mu\nu}\right] = 0
\eeq
underlining the formal similarity of Isometrodynamics to Yang-Mills theories of compact Lie groups.

The $D\times 4$ field equations Eqns.(\ref{58}) clearly display the self coupling of the $A^\nu\,_M$-fields to the $D$ conserved Noether current densities
\beqq \label{63}
& & {\cal J}_\nu\,^M \equiv A^{\mu\,N}\cdot \nabla_N F_{\mu\nu}\,^M
- F_{\mu\nu}\,^N\cdot \nabla_N A^{\mu\,M} \nonumber \\
& & \quad\quad \pa^\nu {\cal J}_\nu\,^M = 0,\:\:M=\sl{1,2},\dots,D  \eeqq
which obey the restrictions on algebra elements $\nabla_M {\cal J}_\nu\,^M = 0$ as expected.

Next we analyze the invariance of the action Eqn.(\ref{56}) under spacetime translations and derive the conserved energy momentum tensor. In the usual way we obtain the canonical energy momentum tensor
\beq \label{64}
T^\mu\,_\nu = -\,\frac{D(D+2)}{\Om_D}\, \Tr_\La \left\{\frac{1}{4}\,\eta^\mu\,_\nu\,F_{\rho\si}\, F^{\rho\si}
- F^{\mu\rho}\, \pa_\nu A_\rho \right\}_g
\eeq
which is conserved $\pa_\mu T^\mu\,_\nu = 0$. As in other gauge field theories this tensor is, however, not gauge invariant. Using the field equations Eqns.(\ref{62}) and the cyclicality of the trace we find
\beq \label{65}
\pa_\rho \Tr_\La \left\{F^{\mu\rho}\, A_\nu \right\}_g = \Tr_\La \left\{F^{\mu\rho}\,\left(\pa_\rho A_\nu + [A_\rho,A_\nu]\right) \right\}_g.
\eeq
Subtracting this total divergence we finally get an improved, conserved and gauge-invariant energy momentum tensor
\beqq \label{66} 
\Th^\mu\,_\nu
&=& T^\mu\,_\nu -\,\frac{D(D+2)}{\Om_D}\, \pa_\rho \Tr_\La \left\{F^{\mu\rho}\, A_\nu \right\}_g \\
&=& -\,\frac{D(D+2)}{\Om_D}\, \Tr_\La \left\{\frac{1}{4}\,\eta^\mu\,_\nu\,F_{\rho\si}\, F^{\rho\si}
- F^{\mu\rho}\, F_{\nu\rho}\right\}_g \nonumber 
\eeqq
which reads in components
\beq \label{67}
\Th^\mu\,_\nu = \intX \La^{D+2} \left\{\frac{1}{4}\,\eta^\mu\,_\nu\, F_{\rho\si}\,^M \cdot F^{\rho\si}\,_M - \,F^{\mu\rho}\,_M \cdot F_{\nu \rho}\,^M \right\}.
\eeq
The corresponding time-independent momentum four-vector reads
\beq \label{68}
{\bf p}_\mu\equiv \int\! d^{\sl 3}x \Th^{\sl 0}\,_\mu
\eeq
and generates the translations in spacetime.

In addition, Isometrodynamics is obviously Lorentz and - at the classical level - scale invariant under the corresponding spacetime and field transformations. We do not display the corresponding conserved currents and charges here.

Let us finally write down the Bianchi identities
\beq \label{69} {\cal D}_\rho^M\,_N F_{\mu\nu}\,^N + \;\mbox{cyclical in}\; \rho,\mu,\nu = 0. \eeq

The equations above define a perfectly consistent classical dynamical system within the Lagrangian framework. Note that in physical observables such as the energy-momentum tensor the "inner" degrees of freedom are integrated over.

As we ultimately aim at quantizing the theory we next turn to develop the Hamiltonian field theory.

\section{Hamiltonian Field Dynamics}
In this section we develop Hamiltonian Isometrodynamics closely following \cite{stw2}. We fix a gauge first choosing Cartesian coordinates along with the Euclidean metric in "inner" space and eliminate the first class constraints related to the remaining gauge degrees of freedom second imposing the axial gauge condition. We give the Hamiltonian ${\cal H}_{ID}$ of the theory in this gauge which will serve in \cite{chw3} as the starting point for quantization. Finally we check the consistency of Hamiltonian Isometrodynamics with the Lagrangian field dynamics in the axial gauge.

Let us use the gauge freedom of Isometrodynamics to choose Cartesian coordinates along with the Euclidean metric in "inner" space. In other words we fix a gauge up to coordinate transformations Eqns.(\ref{42}) which leave the Euclidean metric invariant, i.e. which have an orthogonal Jacobian. Hence, we start with the Lagrangian density Eqn.(\ref{54})
\beq \label{70}
{\cal L}_{ID} (A_\nu\,^M, \pa_\mu A_\nu\,^M, \nabla_N A_\nu\,^M, \La)
= -\frac{\La^2}{4} \, F_{\mu\nu}\,^M \cdot F^{\mu\nu}\,_M,
\eeq
where the $A_\nu\,^M$ are the fundamental variables, where $F_{\mu\nu}\,^M$ is given by Eqn.(\ref{40}) and where the $M$-indices are raised and lowered with $\de_{MN}$.

Next we define the variables $\Pit^\mu\,\!_M$ conjugate to $A_\mu\,^M $ by
\beq \label{71}
\Pit^\mu\,\!_M
\equiv \frac{1}{\La}\, \frac{\pa{\cal L}_{ID}}{\pa(\pa_{\sl 0} A_\mu\,^M)}
= -\,\La\, F^{{\sl 0}\mu}\,_M
\eeq
which are dimensionless in "inner" space. By definition they are elements of the gauge algebra ${\overline{\bf diff}}\,{\bf R}^D$ and fulfil
\beq \label{72}
\nabla^M \Pit^j\,\!_M = 0.
\eeq

As usual we find the two sets of $D$ constraints
\beq \label{73}
\Pit^{\sl 0}\,\!_M = 0
\eeq
and
\beq \label{74}
\pa_k\, \Pit^k\,\!_M + A_k\,^N\cdot \nabla_N\, \Pit^k\,\!_M - \Pit^k\,\!_N\cdot \nabla^N A_{k\,M} = 0
\eeq
which are the field equations Eqn.(\ref{58}) for $\nu = {\sl 0}$. 

To continue let us define the equal-time Poisson bracket of two functionals ${\cal F}_1$ and ${\cal F}_2$ of $A_i\,^M$ and $\Pit^j\,_N$ by
\beq \label{75}
\Big\{{\cal F}_1, {\cal F}_2 \Big\}^P \!\! = \!
\int\! d^3 x \int\! d^D X \La^D \sum_{i, M}
\frac{1}{\La} \left[ \frac{\de{\cal F}_1}{\de A_i\,^M}\,
\frac{\de{\cal F}_2}{\de\, \Pit^i\,_M} -
\frac{\de{\cal F}_1}{\de\, \Pit^i\,_M}\,
\frac{\de{\cal F}_2}{\de A_i\,^M} \right],
\eeq
where all arguments in the denominators are to be taken at $(x_{\sl 0},{\bf x}; X)$ and the factors of $\La$ ensure the correct scaling behaviour of the r.h.s. Note that the brackets above obey the usual rules for commutators and that they respect the constraint on gauge algebra elements Eqn.(\ref{36}).

The Poisson bracket of the two constraints Eqns.(\ref{73}) and (\ref{74})  vanishes because Eqn.(\ref{74}) is independent of $A_{\sl 0}\,^M$. Hence, they are first class. To properly deal with them we fix the remaining gauge degrees of freedom - coordinate transformations which leave the Euclidean metric invariant - by imposing the axial gauge condition
\beq \label{76}
A_{\sl 3}\,\!^M =0
\eeq
now fully fixing the gauge.

The canonical variables of the theory reduce to $A_i\,^M$ and their conjugates $\Pit_j\,\!^N$ 
\beqq \label{77}
\Pit_j\,\!^N &=& \La\, F_{{\sl 0}j}\,^N 
= \La\,\Big(\pa_{\sl 0} A_j\,^N - \pa_j A_{\sl 0}\,^N \nonumber \\
&+& A_{\sl 0}\,^M \cdot \nabla_M A_j\,^N - A_j\,^M \cdot \nabla_M A_{\sl 0}\,^N\Big)
\eeqq
for $i,j = {\sl 1,2}$ only.

$A_{\sl 0}\,^M$ is not an independent variable, but is defined in terms of the canonical variables above and the constraint Eqn.(\ref{74}) which can be recast as
\beqq  \label{78}
\La\,\pa_{\sl 3}\,\!\!^2 A_{\sl 0}\,^M &=& \sum_{i={\sl 1}}^{\sl 2} \left(\pa_i \Pit_i\,\!^M + A_i\,^N\cdot \nabla_N \Pit_i\,\!^M - \Pit_i\,\!^N \cdot \nabla_N A_i\,^M \right) \nonumber \\
&=& \sum_{i={\sl 1}}^{\sl 2} {\cal D}_i^M\,_N \Pit_i\,\!^N,
\eeqq
where we have used $F_{{\sl 30}}\,^M = \pa_{\sl 3} A_{\sl 0}\,^M$. Eqn.(\ref{78}) can be easily solved for $ A_{\sl 0}\,^M$ as a functional of the independent variables $A_i\,^M$ and $\Pit_j\,^N$.

With the constrained $D\times 2$ canonical variables in the axial gauge identified we can write down the corresponding Hamiltonian density ${\cal H}_{ID}$ of the theory
\beqq \label{79}
{\cal H}_{ID}
&\equiv& \La\, \sum_{i={\sl 1}}^{\sl 2} \Pit_{iM}\cdot \pa_{\sl 0}A_i\,^M - {\cal L}_{ID} \nonumber \\
&=& \La\, \sum_{i={\sl 1}}^{\sl 2} \Pit_{iM} \left(\pa_i A_{\sl 0}\,^M
+ A_i\,^N \cdot \nabla_N A_{\sl 0}\,^M 
- A_{\sl 0}\,^N \cdot \nabla_N A_i\,^M \right) \nonumber \\
&+& \frac{1}{2}\, \sum_{i={\sl 1}}^{\sl 2} \Pit_i\,\!^M \cdot \Pit_{iM}  
+ \frac{\La^2}{4}\, \sum_{i,j={\sl 1}}^{\sl 2} F_{ij}\,^M \cdot F_{ij\,M} \\
&+& \frac{\La^2}{2}\, \sum_{i={\sl 1}}^{\sl 2} \pa_{\sl 3}A_i\,^M\cdot 
\pa_{\sl 3}A_{iM} - \frac{\La^2}{2}\, \pa_{\sl 3}A_{\sl 0}\,^M\cdot 
\pa_{\sl 3}A_{{\sl 0}M}, \nonumber
\eeqq
where we have used Eqn.(\ref{74}) to rearrange terms. $A_{\sl 0}\,^M$ is given by Eqn.(\ref{78}) as a functional of the independent canonical variables. We note that from Eqn.(\ref{64}) we find ${\cal H}_{ID} = T^{\sl 0}\,_ {\sl 0}\,_{\mid\,_{ A_{\sl 3}\,^M =\,0}}$ as expected for consistency reasons.

The corresponding time-independent Hamiltonian is given by
\beqq \label{80} 
H_{ID} &=& \int\! d^{\sl 3}x \intX \La^D\, {\cal H}_{ID} \nonumber \\
&=& \int\! d^{\sl 3}x \intX \La^D\, \Bigg\{
\La^2\, \pa_{\sl 3}A_{\sl 0}\,^M \cdot \pa_{\sl 3}A_{{\sl 0}M} \\
&+& \frac{1}{2}\, \sum_{i={\sl 1}}^{\sl 2} \Pit_i\,\!^M \cdot \Pit_{iM}  
+ \frac{\La^2}{4}\, \sum_{i,j={\sl 1}}^{\sl 2} F_{ij}\,^M \cdot F_{ij\,M}
\nonumber \\
&+& \frac{\La^2}{2}\, \sum_{i={\sl 1}}^{\sl 2} \pa_{\sl 3}A_i\,^M\cdot 
\pa_{\sl 3}A_{iM} \Bigg\} \nonumber
\eeqq
with $A_{\sl 0}\,^M$ again given in terms of the independent canonical variables $A_i\,^M$ and $\Pit_j\,^M $ by the non-local expression Eqn.(\ref{78}). In this form the Hamiltonian is explicitly positive definite.

The time evolution of observables in the theory is finally given by the Poisson brackets of a (local) observable $O (x,X)$ with the Hamiltonian
\beq \label{81} 
\pa_{\sl 0}  O (y,Y)= \Big\{H_{ID} , O (y,Y) \Big\}^P_{x_{\sl 0} = y_{\sl 0}}. \eeq
Specifically, the set of equations
\beqq \label{82} 
\pa_{\sl 0} A_i\,^M (y,Y) &=& \left\{H_{ID} , A_i\,^M (y,Y) \right\}^P_{x_{\sl 0} = y_{\sl 0}} \nonumber \\
\pa_{\sl 0} \Pit_j\,\!^N (y,Y) &=& \left\{H_{ID} , \Pit_J\,\!^N (y,Y) \right\}^P_{x_{\sl 0} = y_{\sl 0}}. \eeqq
is equivalent to the Lagrangian field equations Eqn.(\ref{58}) in the axial gauge.

Together, Eqns.(\ref{80}) and (\ref{81}) constitute classical Hamiltonian Isometrodynamics, a perfectly consistent classical field theory for the $A_i\,^M$-fields and their conjugates $\Pit_j\,\!^N$.

\section{Inclusion of "Matter" Fields}
Let us finally comment on the inclusion of "matter" fields. The minimal coupling prescription suggests to couple "matter" by (1) allowing fields to "live" on ${\bf M\/}^{\sl 4}\times {\bf R\/}^{D}$ - adding the necessary additional "inner" degrees of freedom -  and by (2) replacing ordinary derivatives through covariant ones $\pa_\mu\ar D_\mu$ in "matter" Lagrangians as usual. As this prescription involves scalars in "inner" space only and as the volume element $d^D X$ is locally invariant, the metric $g_{MN}$ does not appear in minimally coupled "matter" actions.

Note that this prescription allows for a universal coupling of any "matter" field to the gauge fields of Isometrodynamics treating them as scalars in "inner" space. This universality will form the basis of a potential interpretation of Isometrodynamics as a theory of gravitation.

Technically no fundamentally new difficulties arise and the relevant "matter" terms are simply added to the formulae for both Lagrangian and Hamiltonian Isometrodynamics \cite{stw2}.

\section{Conclusions}

In this paper we have developed Isometrodynamics, the gauge field theory of the group of volume-preserving diffeomorphisms of ${\bf R\/}^{D}$ with unimodular Jacobian, at the classical level, thereby generalizing non-Abelian gauge field theories with a finite number of gauge fields. In contrast to that case, in order to gauge coordinate transformations of an "inner" ${\bf R\/}^{D}$ we had to introduce an uncountably infinite number of gauge fields labeled by $X$, the "inner" coordinates of the fields on which we represent the global and local gauge groups.

This has not brought along fundamental difficulties as far as the definitions of the covariant derivative, the gauge field and the field strength operators are concerned. As the components of these operators are vectors in "inner" space we then introduced a flat metric $g$ on ${\bf R\/}^{D}$ in order to allow for coordinate-invariant contractions of "inner" space indices.

Potentially fundamental difficulties, however, arose in the definition of other  crucial elements of the theory - such as the trace operation in the definition of the action for Isometrodynamics. Tr turned out to be a potentially divergent integral over the non-compact "inner" space ${\bf R\/}^{D}$. Accordingly we have defined the trace operation using the scale parameter $\La$ inherent to the theory as a cut-off and shown that the theories for different $\La$ are in fact related to each other by the global "inner" scale symmetry of the theory.

We then have proposed - with consistent quantization in view - a covariant, minimal Lagrangian for Isometrodynamics. Next, we have derived the field equations and shown their independence of the "inner" metric $g$. Finally we have determined the conserved Noether currents and charges belonging to the inner and spacetime symmetries of the theory. 

The natural framework to consistently deal with gauge fixing, to implement the constraints and to both define Isometrodynamics as a classical field theory and prepare its path integral quantization is the Hamiltonian formalism for which we have derived the theory's Hamiltonian and the corresponding Hamiltonian dynamics through choosing Cartesian coordinates with a Euclidean metric and imposing the axial gauge condition to fully fix the gauge.

The result is a classical field theory formulated on flat four-dimensional Minkowski spacetime which is invariant under local ${\overline{DIFF}}\,{\bf R}^D$ gauge transformations and at most quartic in the fields - a perfect candidate for a renormalizable, asymptotically free quantum field theory.

The quantization and one-loop renormalization of Isometrodynamics as well as its renormalizability are dealt with in a forthcoming paper \cite{chw3}. Separately we will analyze the relevance of this type of theory for a fundamental description of gravity.

\appendix

\section{Notations and Conventions}

Generally, small letters denote spacetime coordinates and parameters, capital letters coordinates and parameters in "inner" space.

Specifically, ({\bf M\/}$^{\sl 4}$,\,$\eta$) denotes $\sl{4}$-dimensional Minkowski spacetime with the Cartesian coordinates $x^\la,y^\mu,z^\nu,\dots\,$ and the spacetime metric $\eta=\mbox{diag}(-1,1,1,1)$. The small Greek indices $\la,\mu,\nu,\dots$ from the middle of the Greek alphabet run over $\sl{0,1,2,3}$. They are raised and lowered with $\eta$, i.e. $x_\mu=\eta_{\mu\nu}\, x^\nu$ etc. and transform covariantly w.r.t. the Lorentz group $SO(\sl{1,3})$. Partial differentiation w.r.t to $x^\mu$ is denoted by $\pa_\mu \equiv \frac{\pa\,\,\,}{\pa x^\mu}$. 
Small Latin indices $i,j,k,\dots$ generally run over the three spatial coordinates $\sl{1,2,3}$ \cite{stw1}.

({\bf R\/}$^{D}$,\,$g$) denotes a $D$-dimensional real vector space with coordinates $X^L, Y^M, Z^N,\dots\,$ and the flat metric $g_{MN}$ with signature $D$. The metric transforms as a contravariant tensor of Rank 2 w.r.t. ${\overline{DIFF}}\,{\bf R}^D$. Because Riem$(g) = 0$ we can always choose global Cartesian coordinates and the Euclidean metric $\de=\mbox{diag}(1,1,\dots,1)$. The capital Latin indices $L,M,N,\dots$ from the middle of the Latin alphabet run over $\sl{1,2},\dots,D$. They are raised and lowered with $g$, i.e. $X_M=g_{MN} X^N$ etc. and transform as vector indices w.r.t. ${\overline{DIFF}}\,{\bf R}^D$. Partial differentiation w.r.t to $X^M$ is denoted by $\nabla_M \equiv \frac{\pa\quad\,}{\pa X^M}$. 

The same lower and upper indices are summed unless indicated otherwise.

\section*{Acknowledgments}

This work is dedicated to the memory of Lochlain O'Raifeartaigh who has taught me the pleasure of doing physics as an intense Socratean dialogue and to my friends Thomas Besmer, Frank Krahe, Lewis Wirshba and Larry Seldon who have kindled that pleasure again and - unknowingly - motivated me to start my search for a gauge theory of the diffeomorphism group all over again.

I owe my warmest thanks to Frank Krahe and Thomas Besmer for all the intensive discussions needed to clarify core elements of the present theory.

\end{document}